\documentclass[structabstract]{aa}
\usepackage{natbib}
\bibpunct{(}{)}{;}{a}{}{,} 
\usepackage{graphicx}
\usepackage{threeparttable}
\usepackage{txfonts}
\def\WD{\mbox{WD\,0836{}+{}201}}
\def\JE{\mbox{J0855{}+{}1640}}

\def\Jon{\mbox{J1300{}+{}5904}}
\def\Rej{\mbox{RE{}\,J~0317-853}}

\def\NGC{NGC\,6819-8}
\newcommand{\Teff}{\hbox{$T_{\rm eff}$}}
\newcommand{\Msolar}{\mbox{\,$M_{\sun}$}}
\newcommand{\Rsolar}{\mbox{\,$R_{\sun}$}}

\begin{document}
   \title{The progenitors of magnetic white dwarfs in open clusters}

   \titlerunning{The progenitors of MWDs in open clusters}


   \author{B. K\"ulebi
          \inst{1,2,3}
	  \and
	J. Kalirai
	\inst{4,5}
	\and
	  S. Jordan
	  \inst{3}
	\and
	  F. Euchner
	  \inst{6}
	  }

	\offprints{B. K\"ulebi, \email{ bkulebi@ieec.cat} }
\institute{Institut de Ci\`encies de l'Espai (CSIC-IEEC), Facultat de
	Ci\`encies, Campus UAB, Torre C5-parell, 2$^{\rm{a}}$ planta, 08193
	Bellaterra, Spain \and
	Institute for Space Studies of Catalonia (ICE), c/Gran Capit\`a 2-4,
	Edif. Nexus 104, 08034 Barcelona, Spain \and 
	Astronomisches Rechen-Institut, Zentrum f\"ur Astronomie der
	Universit\"at Heidelberg, M\"onchhofstr. 12-14, D-69120 Heidelberg,
	Germany \and
	Space Telescope Science Institute, 3700 San Martin Drive,
	Baltimore MD, 21042, USA \and
	Center for Astrophysical Sciences, Johns Hopkins University,
	Baltimore, MD, 21218, USA \and
	Swiss Seismological Service, ETH Zurich, Sonneggstrasse 5,
	8092 Zurich, Switzerland
	}

   \date{Received XXXX, 2012; accepted XXXX, 2012}


  \abstract
   {White dwarfs are the final stages of stellar evolution for most stars
   in the galaxy and magnetic white dwarfs (MWDs) represent at least ten
   percent of the whole sample. According to the fossil-field hypothesis
   magnetic fields are remnants of the previous stages
   of evolution. However, population synthesis calculations are unable to
   reproduce the MWD sample without binary interaction or
   inclusion of a population of progenitor with unobservable small-scale 
   fields.}
   {One necessary ingredient in population synthesis is the
   initial-to-final-mass relation (IFMR) which describes the mass-loss
   processes during the stellar evolution. When white dwarfs are members of open
   clusters, their evolutionary histories can be assessed through the use
   of cluster properties. This enables an independent way of determining
   the mass of their progenitors. The discovery of the magnetic WD\,0836+201 
   in the Praesepe cluster prompted the question whether magnetic fields 
   affect the IFMR. In this work we investigate this suggestion through 
   investigations of all three known MWDs in open clusters.}
   {We assess the cluster membership by correlating the proper-motion 
   of MWDs with the cluster proper-motion and by analyzing the candidates
   spectroscopically with our magnetic model spectra in order to estimate
   the effective temperature and radii. Furthermore, we use mass-radius
   relations and evolutionary models to constrain the histories of the
   probable cluster members.} 
   {We identified SDSS\,J085523.87+164059.0 to be a proper-motion
   member of Praesepe. We also included the data of the formerly
   identified cluster members NGC 6819-8, WD\,0836+201 and estimated
   the mass, cooling age and the progenitor masses of the three
   probable MWD members of open clusters. According to our analysis,
   the newly identified cluster member SDSS\,J085523.87{}+{}164059.0
   is an ultra-massive MWD of mass $1.12\pm0.11\Msolar$.}
   {We increase the sample of MWDs with known progenitor masses to
   ten, with the rest of the data coming from the common proper
   motion binaries. Our investigations show that, when effects
   of the magnetic fields are included in the diagnostics, the
   estimated properties of these cluster MWDs do not show evidence
   for deviations from the IFMR. Furthermore we estimate the
   precision of the magnetic diagnostics which would be necessary
   to determine quantitatively whether magnetism has any effect on
   the mass-loss.} 

   \maketitle

\section{Introduction}
White dwarfs are the final products of stellar evolution for more than
95\% of the stars in our galaxy. One of the most important tools for the
understanding of stellar evolution is the initial-to-final-mass relation
(IFMR), which, links the progenitor mass of a star to the final white dwarf
mass \citep{Weidemann77}. The IFMR is inherently connected to mass-loss
mechanisms \citep[see e.g.][]{Weidemann00}. The understanding of stellar
evolution from the perspective of the IFMR enables us to constrain the
critical mass necessary for Type II supernova explosions. Along with the
initial mass function, this information can be used to estimate the birthrates
and energetics of the supernovae, as well as the birthrates of the SN II
products, the neutron stars \citep{vanderBerghTammann91}. The precise IFMR
also constrains the galactic chemical evolution (i.e. enrichment in the
interstellar medium), which in turn contributes to our understanding of the
star formation efficiency in galaxies \citep{SommervillePrimack99}.

The first cluster IFMR studies were performed by \citet{Weidemann77,
KoesterWeidemann80}. With now about 40 white dwarfs associated with 11
clusters the empirical information on the IFMR has more than doubled
over the last few years \citep[see ][for a recent discussion of cluster
IFMR data]{Kaliraietal09,Casewelletal09,Dobbieetal09, Williamsetal09}.
The increase in the data not only established the support for the already
estimated general trend, in which higher stellar masses yield more massive
white dwarfs, but also resulted in an increase in the scatter. This spread is 
argued to be the result of differences in the host
environments. \citet{Marigo01} argued that one of these effects is metallicity,
for which metal-rich stars undergo heavier mass-loss and yield, on average,
lighter white dwarfs. \citet{Kaliraietal05} was first to show this to be likely
through spectroscopic observations of white dwarfs in NGC 2099, and later
\citet{Kaliraietal07} demonstrated the possibility of this effect in NGC 6791.

Another physical process affecting the final mass of the white dwarf was
proposed to be rotation. The fact that angular momentum acts as an extra
pressure support against gravity inside a degenerate structure has been
known since the works of \citet{Anand65,Roxburgh65,Ostriker66}. Within
the context of stellar evolution simple numerical methods show that this
extra pressure causes the rotating degenerate core to expand and hence
keep the maximum temperature lower than the carbon ignition temperature
\citep{Dominguezetal96}. This lower temperature enables the growth of the
C-O core in mass while avoiding carbon ignition. In this way ultra-massive
(1.2-1.4 \Msolar) white dwarfs can be formed from progenitors with masses
between 6-8 \Msolar. 

Magnetic fields, alongside angular momentum, were also proposed to be
an effective factor of the mass-loss, after the discovery of a magnetic
white dwarf (MWD) \WD\ as a probable member of the Praesepe cluster
\citep[EG59,][]{Claveretal01,Catalanetal08}. Note that it was mislabeled
in both works as explained in \citet{Casewelletal09}. Considering \WD\ as
a Praesepe member, and not taking the magnetic field into consideration,
the estimated white dwarf mass was higher than expected for its progenitor
mass. If magnetism is an important factor affecting the IFMR, this would
have implications for the MWD population. 

MWDs constitute at least 10\% of the white dwarfs if observational
biases are considered \citep{Liebertetal03,Kawkaetal07}. The
current known population of MWDs has been increased considerably
by the Sloan Digital Sky Survey (SDSS) to about 220 objects
\citep{Gaensickeetal02,Schmidtetal03,Vanlandinghametal05,Kulebietal09}. SDSS
also dramatically increased the total known white dwarf population \citep[see
e.g.][for Data Release 7]{Kleinmanetal13} and recent studies indicate that the
number of MWDs in the SDSS could be as large as 521 \citep{Kepleretal13}. The
origin of the magnetic fields of MWDs is still under question. The accepted
picture is the fossil-field hypothesis, where the field strengths are
reminiscent of an earlier stage of stellar evolution. Due to the conservation
of flux, the field strengths are amplified by the contraction of the stellar
core to a white dwarf. For the case of MWDs, chemically peculiar Ap and Bp
stars were proposed as the progenitors \citep{Angeletal81}.

One problem of the fossil-field hypothesis is the relatively massive nature of
the MWDs \citep{Liebert88,VennesKawka08}. While the mean value of the masses of
the MWDs is $0.78\,$\Msolar\ \citep{Kawkaetal07}, the mean mass of the non-magnetic 
white dwarf sample is $0.661\,$\Msolar\ \citep{Gianninasetal11}. The hypothesis 
that this could be a result due to the impact of magnetism on the mass-loss was 
considered by \citet{WickramasingheFerrario05} via population synthesis. Their 
conclusion was that the current number distribution and masses of high-field 
magnetic white dwarfs (HFMWDs, $B\ge10^6\,$G) are not generated by an inclusion of
a modified IFMR, but rather by considering the contribution of about $10$ percent 
of A/B stars with unobservable small scale magnetic fields.

The aim of this work is to analyze the MWDs in clusters with sophisticated
magnetic models, in order to estimate the progenitor masses. This is undertaken
by calculating the masses of the MWDs and their progenitors by using the 
additional information estimated due to their membership in open clusters, 
namely their distances and total evolutionary ages

Up to now, only two MWDs have been discovered as members of open
clusters, one in Praesepe (\WD) and the other one in NGC 6819 \citep[NGC
6819-8,][]{Kaliraietal08}. Neither of them has been analyzed with magnetic
models so far. In order to increase the statistical sample further, we
correlate the properties of SDSS white dwarfs that show evidence of magnetic
fields with the kinematic properties of open clusters in order to determine
the possibility of their cluster membership. While we identified only
SDSS\,J085523.87{}+{}164059.0 (SDSS\,\JE) as a member of an open cluster,
we also analyzed NGC 6819-8. Furthermore, in this work we discuss former
investigations on the influence of magnetism on the IFMR, compare our results
with previous studies, and evaluate the analysis employed.

In Sect.\,\ref{sec:member} we assess the cluster membership of the currently known
SDSS MWDs in the literature through proper-motion information. Afterwards,
we describe our methods to analyze the cluster MWDs in
Sect.\,\ref{sec:analysis} and explain what the relevant uncertainties
are. In Sect.\,\ref{sec:results} we apply our method to cluster MWDs
and in Sect.\,\ref{sec:discussion} we discuss whether we can conclude 
on possible effects of the magnetism on the IFMR 

\label{tab:member}
\begin{table*}
\caption{Photometric and astrometric properties of the DAHs with possible
Praesepe memberships.}
\begin{tabular}{lccccccccc}
\hline
\hline
MWD & RA / (h m s) & Dec. / ($\deg\ '\ ''$) &$r$ / mag &$i$ / mag & $z$ / mag & RA p.m. /
(mas/yr) & Dec. p.m. / (mas/yr)\\
\hline
\WD\ & 08 39 45.56 & +20 00 15.7 & 18.11$\pm$0.01 & 18.36$\pm$0.01 &
18.66$\pm$0.04 &-32.93$\pm$2.87 & -15.98$\pm$2.87 \\
SDSS\,\JE\ & 08 55 23.87 & +16 40 59.0 & 18.80$\pm$0.01 & 19.05$\pm$0.02 &
19.32$\pm$0.08 & -33.14$\pm$2.75 & -14.71$\pm$2.75 \\
\hline
\end{tabular}
\label{tab:photo}
    \begin{tablenotes}
\item \textbf{Notes.} The columns indicate the right ascension (RA), declination (Dec.); the
SDSS photometric magnitudes $r$, $i$, $z$ which are relevant for this work;
and finally the proper-motions. Compare the proper-motions of the objects
with the proper-motions of Praesepe RA p.m.$ = -35.90\pm0.13\,$mas/yr,
Dec. p.m. $= -12.88\pm0.11\,$mas/yr \citep{Kharchenkoetal05}.
   \end{tablenotes}
\end{table*}

\section{Cluster Membership}
\label{sec:member}

In our investigation we compared the astrometric properties of 137 SDSS
hydrogen-rich (DA) MWDs \citep[see ][and references therein]{Kulebietal09}
with the properties of 520 open clusters from the Catalogue of Open Cluster
Data \citep{Kharchenkoetal05}. In order to assess the membership probabilities
of the white dwarfs based on their kinematics, we use the proper-motions
estimated by the SDSS \citep{Munnetal04}. The proper-motions of the MWDs
were acquired from the \texttt{CASJOBS SQL} interface \citep{LiThakar08}.

Our criterion for open cluster membership was through checking whether the
MWD lie within the tidal radius of an open cluster \citep[values taken
from the catalogue of][]{Piskunovetal08} and whether the proper-motion
of the MWD is within $3\,\sigma$ of the cluster proper-motion. As a
result we obtained only one new possible open cluster member SDSS\,\JE\
of the open cluster NGC 2632 (Praesepe), which is a nearby open cluster
that has been the subject of many white dwarf and IFMR investigations
\citep{Luyten62,EggenGreenstein65,Claveretal01}. Additionally with our
calculations we recover the prototypical cluster member MWD, \WD.

Another constraint on possible membership is the position of the white
dwarfs in the color-magnitude diagram (CMD) with respect to the cooling
sequence expected for the cluster \citep[see ][]{Kaliraietal08}. Since
this method involves a prescription for the parameters we want to evaluate,
e.g. deviation from the expected IFMR due to the effect of magnetism, we did
not undertake such an analysis. Rather, the evolutionary constraints on the
cluster membership come from the estimated cooling age of the white dwarf,
i.e. the cooling age cannot be larger than the cluster age.

Finally the membership possibility of both SDSS\,\JE\ and \WD\ can be
assessed through considering the studies made on the WD populations within
the open clusters. Both of these MWDs are possible members of Prasepe. The
current number of white dwarfs in this cluster, confirmed by radial velocity
observations of \citet{Casewelletal09} is eight, including the known magnetic
\WD. compared to numerical simulations of the dynamics of open clusters
\citep{Zwartetal01} the number of white dwarfs found in Praesepe is lower
than expected, especially if compared to the number of observed giants. Hence
it is logical to expect more white dwarfs to be discovered if the search is
extended at least up to the tidal radius, as has been done in our work.

\section{Analysis}
\label{sec:analysis}

One of the necessary ingredients to determine the cooling age of a white
dwarf is the determination of \Teff\ and $\log g$ values. Although this is
undertaken by $\chi^2$ fits to the spectral lines for the non-magnetic white
dwarfs, the situation for the MWDs is more complicated due to the influence
of magnetism on the atomic transitions and the distribution of the magnetic
field strengths over their surface. 

To overcome this difficulty and to analyze the magnetic field geometry of the
MWDs, we use a two-step analysis approach in which first the model spectra for
magnetized atmospheres are calculated for a distribution of magnetic field
vectors with respect to the line of sight and the normal on the surface of
the star \cite[see][]{Jordan92,JordanSchmidt03}. This is repeated for given
\Teff\ and $\log g$ values and a database is generated. 

Later these spectra representing individual vectors could be accessed
from the database and added up in order to calculate the spectrum for a
given geometry. The parameters of the field geometry are determined by an
evolutionary fitting procedure \citep[see][]{Euchneretal02}. 

This approach was used by \cite{Kulebietal09} where the models in
terms of offset dipoles for the single phase of these white dwarfs were
determined. However, for that work the \Teff\ values were determined by
the photometric colors. In this current work, we use the same database and
procedure to determine the \Teff\ value of the MWDs in detail. 

Unfortunately, for MWDs it is not possible to determine the $\log g$ values
due to the difficulties in accounting for the simultaneous impact of Stark
and Zeeman effects on the spectral lines. Moreover, no atomic data exists
for hydrogen in the presence of both magnetic and electric fields for
arbitrary strengths and arbitrary angles between the two fields. Up to now,
only simple cases of parallel electric versus magnetic configurations have
been investigated \citep{Friedrichetal94}. Since in our models only crude
approximations have been used, it is not possible to avoid the systematic
uncertainties, especially in the low-field regime ($\le 5$\,MG) where the
influence of the Stark effect on the spectral lines is strong relative to
the effect of Zeeman splitting. Therefore, we do not determine $\log g$
values from our spectral fits, but rather assume $\log g$ value of 8.0 for
our magnetic model spectra database. This approach have been applied multiple
times in the literature \citep{Gaensickeetal02,Girvenetal10}. 
Additionally, we estimated the error budget introduced by
the undetermined surface gravity through repeating our fitting procedure
for the cases in which $\log g=7.0$ and $\log g=9.0$, and later propagating
the resulting uncertainties.

\begin{figure}
   \centering
   \includegraphics[width=0.5\textwidth]{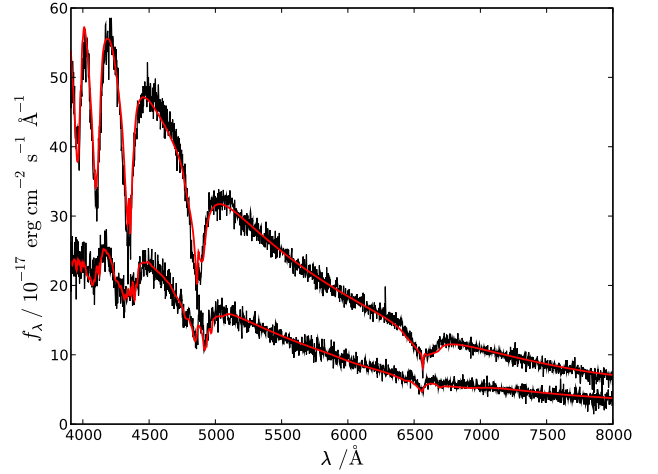}
  \caption{Spectral fits to \WD\ (top) and SDSS\,\JE\ (bottom) determined
  in this work. The parameters are given in Table\,\ref{tab:spectral}. The
  inconsistencies in the \WD\ line fits are due to the simple treatment
  of the relatively large Stark effect in the low-field regime ($B \le
  5$\,MG).}
  \label{fig:spectra}
\end{figure}

The determined parameters are listed in Table\,\ref{tab:spectral}, the errors in
\Teff\ are defined by the spacing of our grid points in our database. 
The observed spectra are compared 
to theoretical models in Fig.\,\ref{fig:spectra}. The fainter 
flux of SDSS\,\JE\ with respect to the flux of \WD\ despite its higher effective 
temperature hints a smaller radius, hence a more massive nature, if we assume
both MWDs to be at the same distance.

\begin{table*}
\centering
\caption{Model fits to the SDSS observations of the objects considered in this
work.}
\begin{tabular}{lccccc}
\hline
\hline
MWD  & Plate-MJD-FiberID & $B_{\rm dipole}$ / MG & $z_{\rm off}$
/ $r_{\rm
WD}$ & $i$ / deg & \Teff\ / K \\
\hline
\WD & 
2277-53705-484  & 
2.83$\pm$0.19 (0.62)& 
-0.26$\pm$0.02 (0.11)&
26$\pm$4 (33)& 
17000$\pm$500 \\
SDSS\,\JE & 
2431-53818-522 & 
12.6$\pm$1.0 (3.9)& 
-0.25$\pm$0.02 (0.16)&
44$\pm$6 (40)& 
20000$\pm$500 \\
\hline
NGC6819-8 & -- & 
10.3$\pm$1.1 (2.4)& 
-0.20$\pm$0.03 (0.09)& 
52$\pm$9 (22)& 
19000$\pm$1000 \\
\hline
\hline
\end{tabular}
\label{tab:spectral}
\begin{tablenotes}
\item \textbf{Notes.} The first two columns indicate the SDSS name of the
object; the plate, Modified Julian Date and fiber ids of the observations; the
remaining columns indicate the model fit parameters for offset dipole models.
The model parameter $i$ refers to the inclination of the magnetic dipole axis
with respect to the line of sight, and the offset is along the axis of the
magnetic field in terms of the stellar radius. The errors in parentheses
refer to the case in which $\log g$ is not fixed to the value of $8.0$,
but is allowed to vary between between the values of $7.0$ and $9.0$.
\end{tablenotes}
\end{table*}

In order to determine the MWD masses we used the synthetic magnitudes
(in the \textit{ugriz} photometric system of SDSS) for carbon-oxygen
(CO) core white-dwarf cooling models with thick hydrogen layer ($M_{\rm
H}/M_\ast=10^{-4}$) masses \citep{Fontaineetal01,HolbergBergeron06}\footnote{
\texttt{http://www.astro.umontreal.ca/$\sim$bergeron/CoolingModels}}; since 
these are non-magnetic models the effect of the magnetism is not included in 
these calculations. 

\begin{figure}
   \centering
   \includegraphics[width=0.5\textwidth]{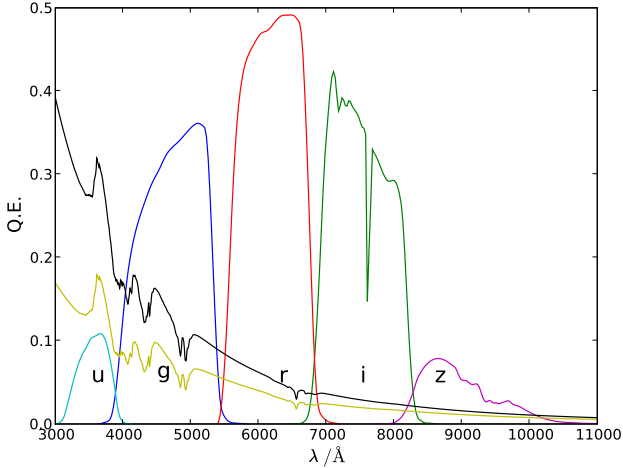}
 \caption[]{The diagram above shows the response curves of the $ugriz$
  photometric pass-bands with 1.3 airmass correction applied, taken from
  the SDSS website\footnotemark, compared to the MWD
  spectrum of two different effective temperatures and same magnetic field geometry
  which for this case is the 12.61\,MG offset dipole model of SDSS\,\JE. The fluxes
  are arbitrarily scaled, but consistent relative to each other.  The effective 
  temperatures are 20\,000\,K (top) and 15\,000\,K (bottom). The $u$, $g$ and $r$ 
  bands cover the wavelengths which are effected by the magnetism in the MWD spectra 
  for field strengths smaller than 50\,MG. However this effect is minor for the 
  $r$ band.}
  \label{fig:ugriz}
\end{figure}

$r$, $i$ and $z$-bands sample mostly the continuum hence these are the
bands least affected by the Zeeman splitting of the Balmer lines (see
Fig.\,\ref{fig:ugriz}). Moreover, in order to minimise the uncertainties of our
analysis we restricted ourselves to the use of $r$, $i$ and $z$. If we compare
the estimated MWD mass from the $i$ band with the masses derived from the $u$
band we obtain a difference as high as 8\%; in the case of masses estimated
from $i$ and $g$ magnitudes, the differences are as small as 1\%. The results
for $z$ and $r$ magnitudes are consistent within 0.05\%. This shows that the
photometric magnitudes that correspond to spectral continuum are consistent
with each other, thus we use the $i$ band in our calculations.

Given the photometric magnitudes and the estimated \Teff\ of an MWD,
we interpolated the synthetic magnitude grids of \citet{HolbergBergeron06}
in order to estimate its mass and cooling age. The systematic errors depend
on observation errors of the photometry and the errors of \Teff\ from
the fits.

Furthermore, we subtracted the estimated cooling ages from the total age
of the open cluster to calculate the progenitor ages of the MWDs. With
this information at hand, estimating the initial masses is rather
straightforward given the stellar evolution models. Hence, we interpolated
the \citet{Bertellietal09} stellar evolution grids with cubic splines to
calculate the progenitor masses given the progenitor lifetime, until the
beginning of the Asymptotic Giant Branch phase.

\begin{table*}
\center
\caption{$\log g$, masses, radii and cooling ages estimated with the photometric
analysis.}
\begin{tabular}{lcccccc}
\hline
\hline
MWD & $\log g$ / dex & $M_{\rm WD}$/ \Msolar\ & radius / 0.01\Rsolar &
$t_{\rm cool}$
/ Myr & $t_{\rm progenitor}$ / Myr & $M_{\rm i}$ / \Msolar \\
\hline \\
\WD &
8.33 $\pm$ 0.08 &
0.82 $\pm$ 0.05 &
1.02 $\pm$ 0.06 &
$221_{-31}^{+37}$ &
$403_{-59}^{+62}$ &
$3.144_{-0.149}^{+0.181}$ \\ \\
SDSS\,\JE &
8.84 $\pm$ 0.22 &
1.12 $\pm$ 0.11 &
0.67 $\pm$ 0.14 &
$391_{-141}^{+170}$ &
$234_{-150}^{+177}$ &
$3.818_{-0.695}^{+1.870}$ \\ \\
\hline \\
NGC6819-8 &
7.77 $\pm$ 0.10 &
0.50 $\pm$ 0.05 &
1.53 $\pm$ 0.76 &
$66_{-10}^{+10}$ &
$2430_{-250}^{+250}$ &
$1.570_{-0.047}^{+0.054}$ \\ \\
\hline
\end{tabular}
\label{tab:results}
\end{table*}

\footnotetext{\texttt{http://www.sdss.org/dr7/}}

\section{Results}
\label{sec:results}

\subsection{\WD}
Previous discussions of the properties of \WD\ are mostly based on the
analysis by \citet{Claveretal01} where the $\pi$ components of the Zeeman
splitted lines were used to determine $\Teff=17\,098\pm350$\,K and
$\log g = 8.32\pm0.05$. Using their values, \citet{Catalanetal08} re-estimated the
mass and the progenitor mass of this object to discuss the effect of the
magnetic field on IFMR. \citet{Casewelletal09} mentioned that the analysis
with non-magnetic model spectra may lead to uncertainties in the temperature
determination. From the SDSS colors ($g-i$, $u-r$) of \WD, they determined
the effective temperature to be about $15\,000\,$K, and argued on
this basis that the \Teff\ and $\log g$ determination from the spectroscopic
non-magnetic model fits are likely to be less robust. With our theoretical
spectra for magnetised white dwarf atmospheres we obtain a more reliable fit
of $\Teff=17\,000\pm500\,$K within the spectral modeling uncertainties, a 
result which is consistent with \citet{Claveretal01}. Since
the estimation of the MWD mass relies on the cluster distance of Praesepe,
additional errors come from the distance errors.

There are multiple distance measurements of Praesepe. Here
we adopt the value of $184.5 \pm 6\,$ pc as the distance to the cluster
center, which was used by \citet{Casewelletal09}. It is the weighted mean
of the Hipparcos-based distance measurement $(m-M)_0=6.24\pm0.12$
\citet{Mermilliodetal97}, the ground-based parallax measurement
$(m-M)_0=6.42\pm0.33$ \citep{GatewooddeJonge94} and the photometric
determination $(m-M)_0=6.33\pm0.04$ \citep{Anetal07}. It is not only a
consistent approach to account for the multiple distance estimations in the
literature, but also advantageous for us since we would like to compare our
estimates to the results of \citet{Casewelletal09}.

For the determination of the progenitor ages, we relied on the cluster age of
$625\pm50\,$ Myr \citep[][from isochrone fits with a metallicity of $Z =
0.024$]{Perrymanetal98}. We have chosen a value of $Z = 0.027$ taken from
\citet{Catalanetal08} but our mass and age determinations do only weakly depend
on the choice of metallicity: If we instead use the lowest recent value for
the metallicity of Praesepe \citep[$Z=0.019$,][]{Claveretal01, Casewelletal09}
the progenitor age becomes only 1\%\ longer, which implies the progenitor mass 
estimates would be 0.03 \Msolar\ smaller, consistent with the estimates from
\citet{Casewelletal09}.

Our results are summarized in Table\,\ref{tab:results}. It should be noted that
the smaller errors of the former works on \WD\ are due to underestimation of
magnetic effects in spectral fits. Moreover, we can only constrain the status
of \WD\ as an outlier to the semi-empirical IFMR up to a limited confidence
interval. Quantifying this depends strongly on the uncertainties in the \Teff\
errors, which are adopted as the difference between the best fit grids in
our models. However, when a greater level of confidence is applied to the
observables used for the parameter estimation, the resulting values for \WD\
entail the semi-empirical IFMR, indicating no evidence for the possible
effect of magnetic fields on the mass-loss. This situation is depicted in
Fig.\,\ref{fig:ifmr}, where we use $3\,\sigma$ errors for the photometric
magnitudes and use $2\,000\,K$ as a more realistic error for the \Teff.

\subsection{SDSS\,\JE}
There are certain caveats for the membership status of SDSS\,\JE\ when the
rest of the Praesepe members are considered, since it is further away from
the cluster center and lies at the limb of the circular region defined by
the tidal radius of the cluster.  Within this radius the cluster members
are expected to have the same kinematic properties as the cluster and high
cluster proper-motions relative to the field stars enable us to discriminate
the members. 

The value for the tidal radius of Praesepe differs in the literature,
and is increasing as the studies become more recent. Starting with
12 pc \citep{Mermilliodetal90}, a higher value of 16 pc was obtained by
\citet{Adamsetal01}, and later an even higher value of $17.1\pm1.2$ pc was
estimated by \citet{Piskunovetal07}. In our work we used the latest value of 
$18.6\pm1.4$ pc value of \citet{Piskunovetal08}.

If we assume that SDSS\,\JE\ lies at 184.5 pc, which is the distance to the core of
the cluster, its angular separation from the cluster center of 4.58$^\circ$
translates into a spatial distance of 14.7 pc. Although numerous values
exist in the literature, the distance of SDSS\,\JE\ from the core puts it safely
within the tidal radius of Praesepe for all determinations except the one
estimated by \citet{Mermilliodetal90}.

Another support for the Praesepe membership of SDSS\,\JE\ is the
determined cooling ages. The value estimated through the \Teff\ value of
$20\,000\,$K (see Table\,\ref{tab:results}) yields cooling ages smaller
than the age of Praesepe. Hence the cluster membership of the SDSS\,\JE\ cannot
be outruled by evolutionary arguments. However, it is still necessary to
carefully consider the spatial position and the distance of SDSS\,\JE\ from the
core of the cluster.

Due to the mass segregation within a cluster as time progresses, relatively
massive stars ($>1\,\Msolar$) concentrate within the core. Since
these stars evolve to white dwarfs, their population is also expected to be
found within the core. This is confirmed for the known white dwarfs
of Praesepe and other clusters. However, there is a selection effect since
spectroscopic investigations specifically aim at the core of the cluster, 
due to the above reasoning. 

However, it is known that the number of observed white dwarfs inside open 
clusters is lower than expected \citep{Weidemann77, Williams02}. The
fact that white dwarfs tend to evaporate from the inner core of an open
cluster was already discussed for the case of the missing white dwarfs in
Hyades \citep{Weidemannetal92}. In this scenario, white dwarfs may leave the
cluster by gaining small velocities ($0.1$ km/s) with respect to the cluster 
center. Recent observations show that there is evidence for this suggestion
for the case of the Hyades open cluster \citep{SchilbachRoeser11}. The 
authors reason that, after the formation of
white dwarfs near the cluster center, due to their lower masses they are
subjected also to mass segregation and move outwards.

Another possibility for gaining a kick velocity of several km/s is due to asymmetric
mass-loss during the AGB phase \citep{Fellhaueretal03}. This was tested
observationally for the globular cluster NGC 6397 and found to be plausible
\citep{Davisetal08}. Since SDSS\,\JE\ is within the tidal radius we did not
consider this case for the following assessment.

Given the cooling age of SDSS\,\JE\ of 390 Myr, a kick velocity of $0.1$
km/s implies the distance travelled since its formation would be about $40$
pc. This value is significantly larger than the tidal radius of Praesepe,
and may explain the distance of SDSS\,\JE\ with respect to the cluster
center. Although a kick velocity explains the distance of SDSS\,\JE\ from
the cluster core, at the same time does not guarantee its radial distance
to be same as the cluster distance. If the MWD has gained a kick velocity
in the radial direction, it may lie considerably behind or in front of the
cluster center.

If the maximum distance travelled is approximately 40 parsecs, and the
distance of SDSS\,\JE\ to the center of the cluster is 14.7 parsecs, then the 
maximum possible unaccounted radial distance equals to 37.2 parsecs. This
is two times the tidal radius if the value of 18.6 parsecs is assumed for 
the value of the tidal radius.

In order to estimate the maximum possible uncertainty, we calculated the progenitor
properties of SDSS\,\JE\ using two tidal radii as the distance error. 
Within these uncertainties the inferred cooling age of
SDSS\,\JE\ is smaller than the age of the cluster, hence its membership
could not be outruled.

Our analysis shows that SDSS\,\JE\ is possibly a part of the rare group of
ultra-massive white dwarfs which have masses higher than $1.1\,\Msolar$.  This
group of objects are proposed to be final products of binary mergers and were
observed numerously in EUV surveys \citep{Bergeronetal91,Marshetal97}. Given
the massive nature and the position on the IFMR of SDSS\,\JE, a merger scenario
could be probable. However, when a larger confidence interval is applied
for calculating the errors of the initial and final mass of this star, these
values would entail the semi-empirical IFMR (see Fig.\,\ref{fig:ifmr}) and
a single star origin cannot be outruled. The nature of this star can only
be understood conclusively, if its \Teff\ errors could be constrained to a
value smaller than 2\,000\,K within $3\,\sigma$ confidence intervals.

\subsection{NGC 6819-8}
\begin{figure}
   \centering
   \includegraphics[width=0.5\textwidth]{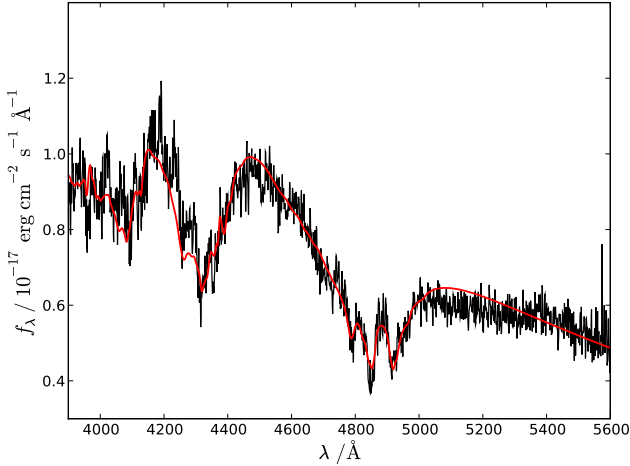}
  \caption{Spectral fit to NGC 6819-8. Fit parameters are given
in Table\,\ref{tab:spectral}. }
  \label{fig:ngc}
\end{figure}

The spectrum of NGC 6819-8 is taken from \citet{Kaliraietal08}, where it was not
analyzed in detail due to the faintness of the star and the low signal-to-noise. 
We have assumed the object to be a member, whereas in \citet{Kaliraietal08} 
the membership assessment was made by comparing the observed $V$ magnitudes 
with the theoretical magnitude from fitting the Balmer lines. The 
comparison indicated that these magnitudes were inconsistent by more 
than $1\,\sigma$, so the white dwarf was assumed not to be a part 
of the cluster.  \citet{Kaliraietal08} used a strict cut to ensure that
only real members are considered, however we cannot employ such a method
for our MWDs due to the complications involved in the modeling of the
photometric magnitudes under the influence of the magnetic fields (see
Sect.\,\ref{sec:analysis}). We assumed the membership of this object, based
solely on the fact that its location agrees with the WD cooling sequence of
the cluster.

Our spectral analysis yielded the \Teff\ and the mass of the white dwarf. Subsequently
we used the same methods used for the Praesepe white dwarfs to estimate the
initial mass of the object. One difference of our analysis was our usage of $V$
magnitudes rather than the $i$ band. This might introduce extra errors since the
$V$ band is influenced by the splitting of hydrogen $\beta$ line. However,
the estimated mass error should be smaller than 1\%, enumerated in
the comparison between estimates with $i$ band which represents the
continuum flux and the $g$ band which covers multiple Balmer lines (see
Fig.\,\ref{fig:ugriz}). For the fits an additional error is introduced
due to the short wavelength interval of the spectrum, which makes it harder
to distinguish between models with different \Teff.

Due to the aforementioned reasons, the main uncertainty in the (final) mass of
the NGC 6819-8 comes from the inaccuracy of the \Teff\ determination. However
the error in the initial mass of the \NGC\ is not affected by the uncertainties
of the spectral modeling since the error of the cluster age dominates over
other uncertainties in the spectral analysis as well as the white dwarf
cooling age itself. This means that the error of the progenitor mass of the
object is comparable to the errors of the progenitor masses of other objects
in the cluster.

The estimation of the cluster age is achieved by theoretical isochrone
fitting, which is dependent precisely on the distance, metallicity and 
reddenning. The distance and metallicity for the cluster NGC 6819 have been
estimated through main sequence fitting by \citet{Kaliraietal01}. In their 
work the adopted reddening of E(B-V) = 0.10-0.14 yielded slightly larger
values ($(m-M)_v = 12.30\pm0.12$) with respect to the previous studies. This
was due to the reddening values being smaller. Using these metallicity,
reddening and distance parameters, the isochrones yield 2.5 Gyr for the age
of the cluster \citep{Kaliraietal01,Kaliraietal08}. It is noted that the
degeneracy of the input parameters might induce a about 10 percent change
in the result --- which is the error adopted for the age in this work ---
however, the parameters are in agreement with the literature and, more
importantly the lower main sequence were reproduced well by their model.

Using the $V$ magnitude of 23.3 for NGC 6819-8, the distance modulus for
NGC 6819 as 12.30$\pm$0.12, and a metallicity of $Z = 0.017$ \citep[all parameters from
][]{Kaliraietal08}, we reached a progenitor mass of $1.57$\Msolar\ which
is similar to the rest of the white dwarfs in the cluster. For comparison,
we have applied our method to get the progenitor mass of the non-magnetic
white dwarfs in NGC 6819, and reproduced the results of \citet{Kaliraietal08}.

Given the uncertainties in the cluster age, the estimated progenitor mass of
\NGC\ is the most precise for the known MWD population. Although the progenitor
mass is relatively low, it is still within the mass range of Ap stars,
which is the progenitor population suggested by the fossil-field hypothesis.

After evaluating the errors and estimating the values of final and progenitor
mass of NGC 6819-8, we end up with values comparable to the rest of the
NGC 6819 white dwarfs. Within the uncertainties we conclude that 
there is no evidence of any effect of magnetism on the mass-loss history 
of NGC 6819-8.

\section{Discussion and Conclusions}
\label{sec:discussion}

\begin{figure}
   \centering
   \includegraphics[width=0.5\textwidth]{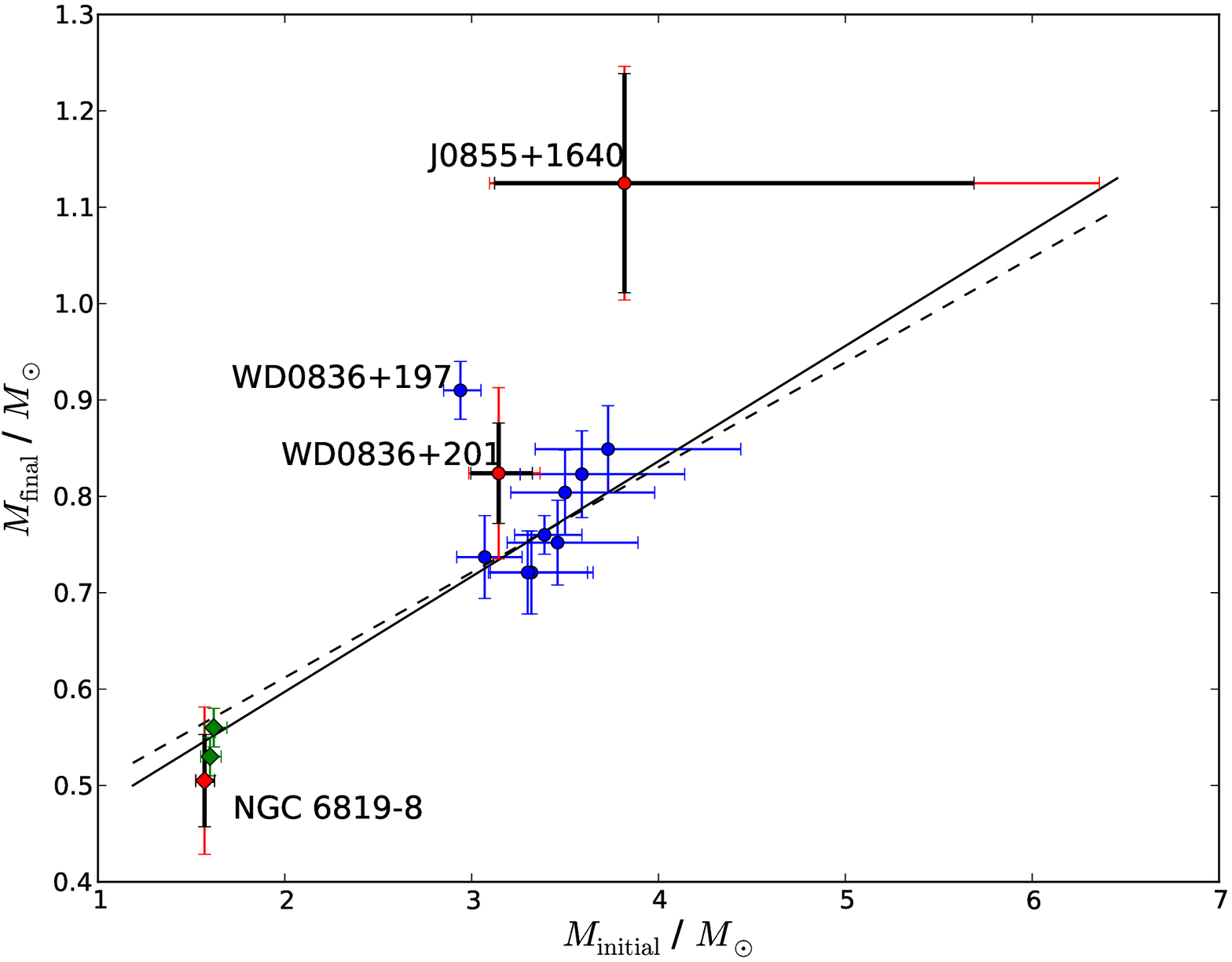}
  \caption{IFMR and the position of the Praesepe white dwarfs
  \citep[circles,][]{Casewelletal09} and NGC 6819 white dwarfs
  \citep[diamonds,][]{Kaliraietal08} are shown in this diagram. The solid
  line is the best fit of \citet{Casewelletal09}, and the dashed line is
  the best fit from \citet{Kaliraietal08} which is sensitive to the low-mass
  end. Red circles denote the positions of the MWDs discussed in this paper,
  and WD\,0836+197 is a known outlier. For the MWDs considered in this work,
  the bold (black) errors correspond to the estimations made with the standard
  procedure as explained in Sect.\,\ref{sec:analysis}, whereas the thin (red)
  error bars correspond to the $3\,\sigma$ error estimations including an
  adopted \Teff\ error of 2\,000\,K.}

  \label{fig:ifmr}
\end{figure}

In this work we investigated the evolutionary histories of MWDs which
are probable members of open clusters. First of all, we correlated the
proper-motions of the known magnetic DA white dwarfs in the SDSS with the
open clusters. Afterwards we estimated the properties of probable cluster
MWDs with our magnetic spectral modeling, and used the cluster properties
to further constrain their evolutionary history. This method was applied
to two currently known cluster MWDs (\WD\ and \NGC) for the first time and
additionally to a newly discovered SDSS\,\JE.

We also compared our results with the the former analyses which have been
undertaken with the use of non-magnetic WD spectral fits. By that, we assessed
the conditions of applicability of non-magnetic analysis for MWDs. Our
results show that SDSS\,\JE\ is probably an ultra-massive MWD and \NGC\
has properties very similar to the rest of NGC 6819 members.

Furthermore, through the use of cluster properties we also estimated the
progenitor masses of these MWDs.  In Fig.\,\ref{fig:ifmr} we compare our
results and the data from Praesepe and NGC 6819 cluster white dwarfs within
the scope of IFMR. Our calculations show that within the uncertainties of the
magnetic analysis, all of the MWDs follow the general trend of the IFMR for
non-magnetic white dwarfs derived from seven open clusters and measurements
from 41 white dwarfs \citep{Casewelletal09}. It should also be noted that
this general trend is not followed by the known outlier WD\,0836+197, for
which neither magnetic field nor rapid rotation was detected; it was proposed
to be a possible radial velocity variable or the product of a blue straggler
\citep{Casewelletal09}. 

The most important diagnostic tool in the estimation of the white dwarf and
the progenitor masses is the determination of the effective temperature. The
current progress in the magnetic spectral analysis allow for the estimating
the best fit for this parameter however the formal determination of the
confidence intervals is still not possible due to the complications caused
by the effect of magnetism on the spectra. Hence in our work we adopt
the distance between the bins in our database as errors. Furthermore,
when we repeat our calculations with $3\,\sigma$ errors for \Teff, we show
that the the estimated progenitor and white dwarf masses would entail the
semi-empirical IFMR values (see Fig.\,\ref{fig:ifmr}). According to these
calculations, to be able to conclusively determine the effect of magnetism
on the mass-loss, the magnetic diagnostics should be able to attain \Teff\
errors smaller than 2\,000\,K within $3\,\sigma$ confidence intervals. In
order to conclusively quantify the confidence levels of complicated $\chi^2$
topographies, a Markov Chain Monte Carlo type of analysis is needed however,
this is beyond the scope of this work.

For the the high-mass white dwarf SDSS\,\JE\ we resolve that it could also be
the result of binary evolution as discussed by \citet{Bergeronetal91} and
\citet{Marshetal97}. In this case it cannot be compared to the IFMR of single
stars. The incidence of magnetism among ultra-massive white dwarfs is rather
high \citep{VennesKawka08}, suggesting a possible relation of the evolution
histories, masses and the magnetic fields.

The membership of SDSS\,\JE\ could be clarified through the determination
of its trigonometric parallax e.g., with Gaia \citep[see
e.g.][]{Torresetal05,Jordan07}. This measurement is also important for 
determining the origin of this star, since our mass determination is based
on the assumption that the star is at the distance of the center of the Praesepe 
open cluster.

For the low mass end, \NGC\ presents similar properties -- if not
a slightly lower progenitor mass -- with respect to the rest of the NGC\,6819 cluster
members. When the position of \NGC\ in the IFMR plot is compared to the
solid line in Fig.\,\ref{fig:ifmr}, which is the best fit of the IFMR
for the low-mass end \citep{Kaliraietal08}, no considerable difference can
be observed within the error bars.

One should also note that there is considerable scatter for the estimated
initial masses smaller than about $3\,\Msolar$ which was also observed for NGC
2099 \citep[M 37,][]{Kaliraietal05}. It was proposed that the theoretical
uncertainties of mass-loss in this regime might also be the reason for this
discrepancy. At this stage it is difficult to distinguish between the observed
scatter in the low mass end due to other physical mechanisms (i.e. metallicity),
and the effect of magnetism. 

It is also possible to test the IFMR using double white dwarfs in wide binary
systems, since these systems are assumed to have evolved
without interacting. The number of known MWDs in WD-WD wide binaries is
six, namely \Rej\ \citep{Barstowetal95,Ferrarioetal97}, SDSS\,\Jon\
\citep{Girvenetal10} and recently SDSS\,J092646.88{}+{}132134.5
(SDSS\,J0926+1321), SDSS\,J150746.80{}+{}520958.0
\citep[SDSS\,J1507+5209,][]{Dobbieetal12},
SDSS\,J074853.07{}+{}302543.5 and SDSS\,J150813.24{}+{}394504.9
\citep{Dobbieetal13}. Their evolutionary status presents different challenges.

The ultra-massive and hot nature of \Rej\ with respect
to its non-magnetic counterpart turns out to be problematic for a single
star. This problem of ``age dilemma'' was resolved by invoking a scenario
where the system initially consists of three components and \Rej\ is the
result of a close binary merger \citep{Ferrarioetal97}. A similar evolutionary
dilemma was also observed for SDSS\,J1507{}+{}5209, where the MWD component
of the binary is relatively hot and young compared to the non-magnetic
one \citep{Dobbieetal12}, hence a triple system was also suggested as an
explanation. The remaining MWD in wide WD-WD pairs do not show similar problems.

A similar type of analysis can be undertaken with common proper motion (CPM)
binaries consisting a white dwarf and a FGK type star, where the total age
and the metallicity of the white dwarf can be assessed by analyzing its
companion. \citet{Catalanetal08cpm} undertook the first of these studies
in which one of the objects (40\,Eri\,B, WD\,0413-017) is an MWD. Their
observational data show dispersion with respect to the semi-empirical IFMR,
however whether the cause of this dispersion is magnetism and/or rotation
has not been discussed in detail. It should be noted that the magnetic
field strength of 40 Eri B \citep[$\approx 2.3$\,kG,][]{Fabrikaetal00}
is significantly lower than the magnetic field of MWDs in this work ($>1$\,MG).

Currently there are ten MWDs with determined white dwarf and progenitor
masses. Not considering the two objects \Rej\ and SDSS\,J1507{}+{}5209 which
are believed to have formed by mergers, it has been suggested that magnetism
could be a factor in modifying the mass-loss based on the discovery of \WD\
\citep{Claveretal01}. However our analysis, which also entails \WD, shows
that when the effect of magnetism on the observed spectrum is included in the
uncertainties, the deviation of the properties of MWDs from the semi-empirical
IFMR disappears. The conclusive test for the extent of the dispersion caused
in the IFMR by the magnetic fields will be possible through better magnetic
diagnostics and more precise distance measurements.

\begin{acknowledgements}
     We thank Andreas Just, Sigfried R\"oser and Elena Schilbach for useful
discussions concerning the kinematics of open cluster members. We also thank
Silvia Catal\'an for her useful comments on the manuscript. This work was
partly supported by the DLR under grant 50 OR 0802. BK acknowledges support by the
MICINN grant AYA08-1839/ESP, by the ESF EUROCORES Program EuroGENESIS (MICINN
grant EUI2009-04170), by the 2009SGR315 of the Generalitat de Catalunya and
EU-FEDER funds.

\end{acknowledgements}

\bibliographystyle{aa}
\bibliography{./aa18842-12.bbl}

\end{document}